\documentclass[%showpacs,
showkeys,12pt,
preprint,preprintnumbers,nofootinbib,
groupedaddress,superscriptaddress,amsmath,amssymb]{revtex4}
\usepackage{graphicx}% Include figure files
\usepackage{dcolumn}% Align table columns on decimal point
\usepackage{bm}% bold math
\usepackage{amssymb}
\usepackage{amsmath}
\usepackage{epsfig}    
\usepackage{color}
\usepackage{slashed}
\usepackage{hhline}
\def\be{\begin{equation}}
\def\ee{\end{equation}}
\newcommand{\bea}{\begin{eqnarray}}
\newcommand{\eea}{\end{eqnarray}}
\newcommand{\nn}{\nonumber}

\numberwithin{equation}{section}

\begin{document}

 \begin{flushright} {CTP-SCU/2020040}, APCTP Pre2020-034  \end{flushright}

%%%%%%%%%
\title{Vector dark matter from a gauged $SU(2)$ symmetry}

\author{Takaaki Nomura}
\email{nomura@scu.edu.cn}
\affiliation{College of Physics, Sichuan University, Chengdu 610065, China}

\author{Hiroshi Okada}
\email{hiroshi.okada@apctp.org}
\affiliation{Asia Pacific Center for Theoretical Physics, Pohang 37673, Republic of Korea}
\affiliation{Department of Physics, Pohang University of Science and Technology, Pohang 37673, Republic of Korea}

\author{Seokhoon Yun}
\email{syun@pd.infn.it}
\affiliation{School of Physics, Korea Institute for Advanced Study, Seoul 02455,  South Korea}
\affiliation{Dipartimento di Fisica e Astronomia, Universit\`a degli Studi di Padova, Via Marzolo 8, 35131 Padova, Italy}
%\affiliation{Istituto Nazionale di Fisica Nucleare (INFN), Sezione di Padova, Via Marzolo 8, 35131 Padova, Italy}

\date{\today}

\begin{abstract}
We propose a scenario of dark sector described by a hidden $SU(2)$ gauge symmetry which is broken by a vacuum expectation value(VEV) of a scalar multiplet.
We discuss a general mass relation among $SU(2)$ gauge bosons after spontaneous symmetry breaking 
which is determined by weight of gauge group representation associated with a scalar multiplet developing VEV.  
Then a model with quintet and triplet scalar fields is discussed in which hidden gauge boson can be dark matter(DM) stabilized by remnant discrete $Z_2$ symmetry and resonant dark matter annihilation is realized by mass relation between DM and mediator.
We estimate relic density and spin independent DM-nucleon scattering cross section searching for allowed parameter region. 
 \end{abstract}
\maketitle
\section{Introduction}

Explanation of dark matter(DM) nature is one of the outstanding issues which requires physics beyond the standard model(SM).
One promising candidate of DM is a stable particle weakly interacting with the SM particles where its relic density in current Universe can be explained by thermal production.
Interactions among DM and the SM particles have been tested by direct detection, indirect detection and collider experiments.
However we have no clear signal of DM in such experiments indicating that interaction among DM and the SM particle would be sufficiently weak to escape these searches.
 
In describing DM physics, one attractive scenario is introduction of dark sector that is described by hidden gauge symmetry as the SM sector is also described by gauge symmetry;
here we indicate hidden gauge symmetry under which the SM fields are neutral.  
In such a scenario we can explain stability of DM by a remnant of hidden gauge symmetry after spontaneous symmetry breaking~\cite{Krauss:1988zc, Ko:2018qxz}.
In particular a case of non-Abelian group is interesting since an unbroken discrete symmetry can be naturally preserved; 
in Abelian case we might need to tune charge assignment since its choice is more arbitrary compared to non-Abelian case.
Moreover hidden gauge boson(s) can be DM candidate depending on a structure of remnant symmetry.
Here we focus on the simplest non-Abelian gauge group $SU(2)$ as hidden gauge symmetry.
In fact, various approaches applying a hidden $SU(2)$ gauge symmetry have been discussed in literatures, for examples, a remaining $Z_{2,3,4}$ symmetry with a quadruplet(quintet) in ref.~\cite{Chiang:2013kqa, Chen:2015nea,Chen:2015dea,Chen:2015cqa,Ko:2020qlt}, $Z_2 \times Z'_2$ symmetry~\cite{Gross:2015cwa}, a custodial symmetry in refs.~\cite{Boehm:2014bia, Hambye:2008bq}, an unbroken $U(1)$ from $SU(2)$ in refs.~\cite{Baek:2013dwa,Khoze:2014woa,Daido:2019tbm}, a model adding hidden $U(1)_h$~\cite{Davoudiasl:2013jma}, a model with classical scale invariance~\cite{Karam:2015jta}, Baryogengesis~\cite{Hall:2019ank} and electroweak phase transition~\cite{Ghosh:2020ipy}.

In this paper, we consider a scenario of dark sector described by hidden $SU(2)$ gauge symmetry which is broken by a vacuum expectation value(VEV) of scalar multiplet(s).
Then we show general mass relation among $SU(2)$ gauge bosons after spontaneous symmetry breaking which is determined by weight of gauge group representation associated with a scalar multiplet developing VEV.
It is found that hidden gauge bosons can be DM and mediator $Z'$ with mass ratio of $1:2$ realizing resonant DM annihilation when $SU(2)$ is broken by scalar quintet VEV.
In our scenario $Z'$ from hidden $SU(2)$ can interact with SM particles via kinetic mixing coming from higher dimensional operator with triplet VEV.
We then discuss relic density and DM-nucleon scattering cross section in such a scenario.

This paper is organized as follows.
In Sec.~II, we show general relation among $SU(2)$ gauge bosons after spontaneous symmetry braking.
In Sec.~III we consider a model realizing resonant annihilation of DM and estimate relic density of DM and spin independent nucleon-DM scattering cross section.
We then conclude and discuss in Sec.~IV.
%\newpage

%%%%%%%%%%%%%%%%%%%%%%%%%%%%%%%%%%%%%
%\section{The Model}
%\subsection{Model setup}

\section{General aspect }
In this section we discuss $SU(2)$ gauge symmetry which is spontaneously broken by a VEV of a scalar field, 
and show relation among massive gauge boson masses after symmetry breaking in general way. 

Let us start from the definition of a scalar field $\Phi$ belonging to $SU(2)$ multiplet whose components are described by $\ell,m$, where numbers $\ell$ and $m$ respectively correspond to the highest weight and weight (eigenvalue of diagonal generator $T_3$). Then, we symbolize it as 
\begin{align}
\Phi\equiv \left[\phi_{\ell,\ell}^{(2\ell)}, \cdots, \phi_{\ell,m+1}^{(\ell+m+1)}, \phi_{\ell,m}^{(\ell+m)}, \phi_{\ell,m-1}^{(\ell+m-1)},\cdots , \phi_{\ell,-\ell}^{(0)} \right]^T,
\end{align}
where the number of component is $2\ell+1$; $-\ell\le m\le \ell$, which corresponds to the dimension of representation of our gauged $SU(2)$ symmetry, and the upper indices represent an order of components from $0$ to $2 \ell$. 
%%%

Here, we show several examples in the $SU(2)_L \times U(1)_Y$ case.
If $\Phi$ is Higgs field in the SM; $\ell=\frac12$ and hypercharge $Y=\frac12$, one writes it to $\Phi=\left[\phi_{\frac12,\frac12}^{(1)},\phi_{\frac12,-\frac12}^{(0)} \right]^T$.
Here upper indices coincide with electric charge since  it is defined by $Q=T_3+Y$ and $Y = \ell$; each of the charges is 1 and 0.
%%%
As another example,
let us consider the Higgs triplet with $\ell=1$ and $Y=1$, which is often used to explain the active Majorana neutrino mass matrix via type-II seesaw mechanism.
In this case, we have $\Phi=\left[\phi_{1,1}^{(2)} ,\phi_{1,0}^{(1)},\phi_{1,-1}^{(0)} \right]^T$.
The upper indices coincide with electric charge again due to $Y=\ell$ and then each of the charges is found to be  2, 1 and 0 that are consistent with the component of Higgs triplet.
Notice here that we consider $SU(2)$ is a hidden symmetry, thus nonzero VEV can be found in the components with nonzero charge. 
% We expect that only the component with zero charge can have nonzero vacuum expectation value.
 
Next, let us consider the kinetic term of $\Phi_{\ell,m}$, which is written by 
\begin{align}
&{\cal L}_{kin}= (D_\mu\Phi_{\ell,m})^\dag (D^\mu\Phi_{\ell,m}),\\
%%%
&D_\mu =\partial_\mu -i\frac{g_X}{\sqrt2}(T^+ X^+_\mu + T^- X^-_\mu)-i g_X T_3 X_\mu^3,
\end{align}
where $X^{\pm}_\mu\equiv (X^1_\mu\mp iX^2_\mu)/\sqrt2$ and the following relations are found
\begin{align}
&T^+ \left[ \phi_{\ell,\ell}^{(2\ell)}, \cdots, \phi_{\ell,m+1}^{(\ell+m+1)}, \phi_{\ell,m}^{(\ell+m)}, \phi_{\ell,m-1}^{(\ell+m-1)},\cdots ,\phi_{\ell,-\ell}^{(0)} \right]^T\nn\\
& = %\sqrt{(\ell-m)(\ell+m+1)}  
 \left[C_{\ell,\ell-1}^+\phi_{\ell,\ell-1}^{(2\ell-1)}, \cdots, C_{\ell,m}^+\phi_{\ell,m}^{(\ell+m)}, C_{\ell,m-1}^+\phi_{\ell,m-1}^{(\ell+m-1)}, C_{\ell,m-2}^+ \phi_{\ell,m-2}^{(\ell+m-2)},\cdots , 0 \right]^T ,\\
%&T^+\phi_{\ell,m}^{(\ell+m)}=\sqrt{(\ell-m)(\ell+m+1)} \phi_{\ell,m}^{(\ell+m)},\\
&T^- \left[ \phi_{\ell,\ell}^{(2\ell)}, \cdots, \phi_{\ell,m+1}^{(\ell + m+1)}, \phi_{\ell,m}^{(\ell+m)}, \phi_{\ell,m-1}^{(\ell+m-1)},\cdots ,\phi_{\ell,-\ell}^{(0)} \right]^T\nn\\
& = %\sqrt{(\ell+m)(\ell-m+1)}  
\left[0, \cdots, C_{\ell,m+2}^-\phi_{\ell,m+2}^{(\ell + m+2)}, C_{\ell,m+1}^- \phi_{\ell,m+1}^{(\ell+m+1)}, C_{\ell,m}^-\phi_{\ell,m}^{(\ell+m)}, \cdots , C_{\ell,-\ell+1}^-\phi_{\ell,-\ell + 1}^{(1)} \right]^T ,\\
 %&T^- \phi_{\ell,m}^{(\ell+m)}=\sqrt{(\ell+m)(\ell-m+1)} \phi_{\ell,m}^{(\ell+m)},\\
%%%
&T_3 \left[\phi_{\ell,\ell}^{(2\ell)}, \cdots, \phi_{\ell,m+1}^{(\ell + m+1)}, \phi_{\ell,m}^{(\ell+m)}, \phi_{\ell,m-1}^{(\ell+m-1)},\cdots ,\phi_{\ell,-\ell}^{(0)} \right]^T\nn\\
& =
 \left[\ell \phi_{\ell,\ell}^{(2\ell)} \cdots, (m+1)\phi_{\ell,m+1}^{(\ell + m+1)}, m\phi_{\ell,m}^{(\ell+m)}, (m-1)\phi_{\ell,m-1}^{(\ell+m-1)},\cdots ,-\ell \phi_{\ell,-\ell}^{(0)} \right]^T , %&T_3 \phi_{\ell,m}^{(\ell+m)}=m \phi_{\ell,m}^{(\ell+m)}.
\end{align}
where we defined $C_{\ell,m}^+\equiv \sqrt{(\ell-m)(\ell+m+1)}$ and  $C_{\ell,m}^-\equiv \sqrt{(\ell+m)(\ell-m+1)}$.

Then, we find the following mass terms for vector bosons when only $\langle\phi_{\ell,m}^{(\ell+m)}\rangle=v_\Phi/\sqrt2\neq0$ 
%and the others are vanishing VEVs;
\begin{align}
& \frac{g_X^2}{4} \left[  (C_{\ell,m}^+)^2 +(C_{\ell,m}^-)^2 \right] v_\Phi^2 X^+_\mu X^{-\mu}
%\frac{g_x^2}{2} (C_{\ell,m-1}^+)^2 v^2 X^+_\mu X^{-\mu}+\frac{g_x^2}{2} (C_{\ell,m+1}^-)^2  v^2 X^+_\mu X^{-\mu}
+ \frac{g_X^2}{2}  m^2   v_\Phi^2 X^3_\mu X^{3\mu}\nn\\
%%%
&\equiv m_{X^\pm}^2 X^+_\mu X^{-\mu} + \frac12 m_{X^3}^2 X^3_\mu X^{3\mu}.
\end{align}
Taking the mass ratio between $X^\pm_\mu$ and $X^3_\mu$, one finds 
\begin{align}
\frac{m_{X^3}^2}{m_{X^\pm}^2} = \frac{4 m^2}{(C_{\ell,m}^+)^2 + (C_{\ell,m}^-)^2} 
= \frac{4 m^2}{(\ell-m)(\ell+m+1) + (\ell+m)(\ell-m+1) }.
\end{align}
Here we consider only one scalar field develops a VEV but it is straitforward to extend more general cases in which multiple scalar fields and/or more than one components in a multiplet develop non-zero VEVs.  

In case of $\ell=|m|=2$, especially, one finds the mass ratio of gauge bosons as follows:
\begin{align}
\frac{m_{X^3}^2}{m_{X^\pm}^2} =4.\label{eq:mr}
\end{align}
Here $\ell=2$ case corresponds to a quintet scalar field $\Phi [\phi^{(4)}_{2,2},\phi^{(3)}_{2,1},\phi^{(2)}_{2,0},\phi^{(1)}_{2,-1},\phi^{(0)}_{2,-2}]^T$,
where only $\phi^{(4)}_{2,2}$ has non-vanishing VEV~\footnote{This relation is pointed out in ref.~\cite{Chiang:2013kqa}}.
Remarkably Eq.~(\ref{eq:mr}) indicates that $X^\pm_\mu$ annihilates into pairs of SM particles around the resonant point when $X^\pm_\mu$ is DM and $X^3_\mu$ is mediator decaying into SM fields.
Note also that a component with even and odd $m$ have respectively $Z_2$ parity even and odd where this remnant $Z_2$ can stabilize $X^\pm$ since it has $m = \pm 1$.
We discuss a model with a quintet scalar field realizing resonant DM annihilation scenario in next section. 
%%%

\section{A model with resonant dark matter annihilation}

In this section we consider a model with hidden $SU(2)_X$ gauge symmetry introducing 
real quintet and triplet scalar fields denoted by $\Phi$ and $\Phi' = \left[\phi'^{(2)}_{1,1},\phi'^{(1)}_{1,0},\phi'^{(0)}_{1,-1} \right]^T$, respectively.
In this setup resonant DM annihilation is realized by mass relation between DM and mediator gauge boson, 
and interaction between dark sector and the SM sector is obtained via kinetic mixing between $SU(2)_X$ and $U(1)_Y$ gauge fields.
In our model scalar potential is written by
\begin{align}
V  = &  M_\Phi^2 \Phi^2 + M_{\Phi'}^2 \Phi'^2 + \mu_1 \Phi \Phi'^2 + \mu_2 \Phi' \Phi^2 + \sum_i \lambda_i (\Phi^4)_i + \sum_a \lambda_a (\Phi'^4)_a \nonumber \\
& + \lambda_{H\Phi} |H|^2 \Phi^2 + \lambda_{H\Phi'} |H|^2 \Phi'^2 + V(H),
\end{align}
where $H$ is the SM Higgs, $V(H)$ is scalar potential including only $H$, each term is contracted to be $SU(2)_X$ singlet implicitly, subscript $i(a)$ distinguishes different contraction of $SU(2)_X$ representation indices.
In our analysis we do not discuss details of scalar potential and assume our VEV alignments are obtained by turning parameters in potential.

\subsection{Mass spectrum and relevant interactions}

Firstly, we discuss our gauge symmetry breaking pattern when $\phi^{(4)}_{2,2}$ and $\phi'^{(0)}_{1,0}$ develop non-zero VEVs. 
The quintet VEV $\langle\Phi\rangle$ is invariant under the unitary transformation of $U_{T_3}=e^{i\pi T_3}$;
$U_{T_3} \langle\Phi\rangle=\langle\Phi\rangle$, where $T_3\equiv{\rm diag}(2,1,0,-1,-2)$. Then, each of the components of $\Phi$ and $\Phi'$ has parity
odd and even; ${\rm Parity} (\phi^{(4)}_{2,2},\phi^{(3)}_{2,1},\phi^{(2)}_{2,0},\phi^{(1)}_{2,-1},\phi^{(0)}_{2,-2})=(+,-,+,-,+)$ and ${\rm Parity}(\phi'^{(2)}_{1,1},\phi'^{(1)}_{1,0},\phi'^{(0)}_{1,-1})=(-,+,-)$ under $U_{T_3}$ transformation.
It suggests that we have a remnant $Z_2$ symmetry that plays a role in stabilizing the DM candidate with odd parity as discussed below.
Note also that if we introduce a field having half integer $\ell$ (without developing a VEV), a remnant symmetry becomes $Z_4$; for example a ($\ell=1/2$) doublet field $\eta = (\eta_1, \eta_2)^T$ transform as $U_{T_3} \eta = (i \eta_1, -i \eta_2)^T$.

%%%
Next step is to find how $X^3_\mu$  interacts with the SM fields.
In order to realize the interaction, we introduce an $SU(2)_X$ triplet scalar field $\varphi$ which is given by 
\begin{equation}
\varphi = \varphi_a \frac{\sigma^a}{2} = \begin{pmatrix} \frac{1}{2} \phi'^{(1)}_{1,0} & \frac{1}{\sqrt2} \phi'^{(2)}_{1,1} \\ \frac{1}{\sqrt2} \phi'^{(0)}_{1,-1} & -\frac{1}{2}\phi'^{(1)}_{1,0} \end{pmatrix},
\end{equation}
where the elements in the most right side correspond to our standard form~\footnote{The relations between $\varphi_a$ and components in our standard form; $\phi'^{(\ell+m)}_{1,m}$, are given by $\phi'^{(2)}_{1,1} = (\varphi_1 - i \varphi_2)/\sqrt{2}$, 
$\phi'^{(0)}_{1,-1} = (\varphi_1 - i \varphi_2)/\sqrt{2}$ and $\phi'^{(1)}_{1,0} = \varphi_3$. }.
Then we can write the following term:
The relevant term for these kinetic mixing is dim-5 operator~\cite{Arguelles:2016ney, Ko:2020qlt}:
\begin{align}
\mathcal{L}_{XB} = & \frac{C_\varphi}{\Lambda} X^a_{\mu \nu} \tilde{B}^{\mu \nu} \varphi^a  
\end{align}
where $\Lambda$ indicates the cut off scale and $X^a_{\mu \nu} = \partial_\mu X^a_\nu -\partial_\nu X^a_\mu + \epsilon^{abc} g_X X^b_\mu X^c_\mu$ and $\tilde{B}^{\mu \nu}$ are the gauge field strength for $SU(2)_X$ and $U(1)_Y$, respectively ($g_X$ and $\epsilon^{abc}$ are $SU(2)_X$ gauge coupling and anti-symmetric tensor).
After $\varphi$ developing VEV, 
\begin{equation}
\langle \varphi \rangle = \begin{pmatrix} \frac{1}{2 \sqrt{2}} v_\varphi & 0 \\ 0 & -\frac{1}{2 \sqrt{2}} v_\varphi \end{pmatrix}, 
\end{equation}
we obtain the following kinetic mixing term: 
\begin{equation}
\mathcal{L}_{\rm KM} = - \frac{1}{2} \sin {\chi} X^3_{\mu \nu} \tilde{B}^{\mu \nu},
\label{eq:kinetic_mixing}
\end{equation}
where we define $\sin {\chi} \equiv \sqrt{2} C_{\varphi} v_\varphi/\Lambda$ as a new kinetic mixing parameter.
%{\color{red} I REALIZED THAT $\chi_a$ , $\chi_2$ here as kinetic mixings could be confused with $\chi_1$ and $\chi_2$ in Eq. (II.1). Could you use different notations for either ones ? See Eq. (III.8) below for example.} {\color{magenta} kinetic mixing parameters $\chi_{1,3}$ are changed to $\delta_{1,3}$}

\begin{figure}[t]
\begin{center}
\includegraphics[width=80mm]{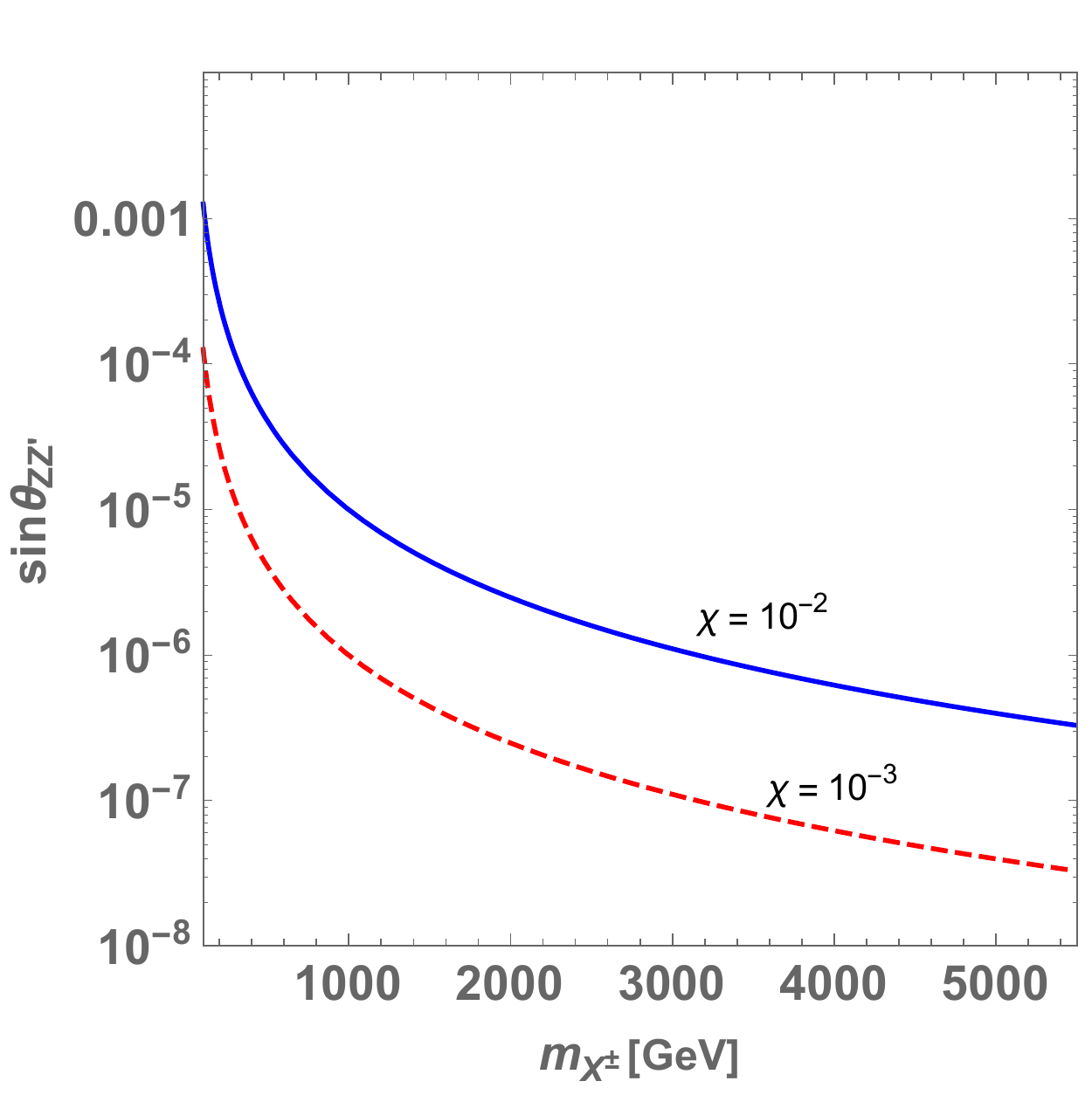} 
\caption{$\sin \theta_{ZZ'}$ as a function of $m_{X^\pm}$ for $\chi =10^{-2}$ and $10^{-3}$ where we fix $2v_\varphi^2/v_\Phi^2 = 0.01$.} 
 \label{fig:mixing}
\end{center}\end{figure}

The kinetic terms for $X^{3}_\mu$ and $\tilde{B}_\mu$ can be diagonalized by the following transformations (we neglect refractive phenomena contributed by a forwardly scattering off the thermal bath, i.e. plasma effect~\cite{Braaten:1993jw,An:2013yfc,Redondo:2013lna,Redondo:2008ec,Hardy:2016kme,Hong:2020bxo}): 
\begin{align}
& \tilde{B}_\mu = B_\mu - \tan { \chi} \tilde{X}^3_\mu , \\
& X^3_\mu = \frac{1}{\cos { \chi} } \tilde{X}^3_\mu.
\end{align}
%where ${\delta}$ is defined as $\sin { \delta} \equiv - \tan {\delta_1} \tan { \delta_3}$.
In our analysis, we take a limit of ${\chi} \ll 1$  and gauge fields are approximately written by
\begin{equation}
\tilde{B}_\mu \simeq B_\mu - { \chi} X^3_\mu ,  \quad X^3_\mu \simeq \tilde{X}^3_\mu.
\end{equation}
Here we denote dark gauge bosons associated with $X^{1,2,3}_\mu$ field as $X_{1,2,3}$ henceforth.
After quintet and triplet scalar fields develop nonzero VEVs, we obtain mass terms for $SU(2)_X$ gauge fields and SM Z boson field  
including kinetic mixing effect such that
\begin{align}
& L_{M} = \frac{1}{2} m_{Z_{SM}}^2 \tilde{Z}_\mu \tilde{Z} + m_{Z_{SM}}^2  \chi \sin \theta_W \tilde Z_{\mu} X^{3 \mu} + \frac{1}{2} m_{X^3}^2 X^{3}_\mu X^{3 \mu} + m_{X^\pm}^2 X^+_\mu X^{- \mu}, \\
& m_{Z_{SM}}^2 = \frac{v^2}{4} (g^2+g_B^2), \quad m_{X^3}^2 = 4 g_X^2 v_\Phi^2, \quad m_{X^\pm}^2 = g_X^2 v_\Phi^2 \left(1+\frac{ v_\varphi^2}{v_\Phi^2}\right),
\end{align}
where $\tilde Z$ is $Z$ boson field in the SM, $g$ and $g_B$ are gauge couplings of $SU(2)$ and $U(1)_B$, and $v$ is the SM Higgs VEV.
Diagonalizing $\tilde Z$ and $X^3$ mass terms, mass eigenstate and mixing angles are given by
\begin{align}
& m_{Z, Z'}^2 = \frac{1}{2} (m_{X^3}^2 + m_{Z_{SM}}^2 ) \mp \frac{1}{2} \sqrt{(m_{X^3}^2- m_{Z_{SM}}^2)^2 + 4 \chi^2 \sin^2 \theta_W m_{Z_{SM}}^4}, \\
& \tan 2 \theta_{ZZ'} = \frac{2 \sin \theta_W \chi m_{Z_{SM}}^2}{m_{Z_{SM}}^2 - m_{X^3}^2},
\end{align}
where we approximate $m_Z \simeq m_{Z_{SM}}$ and $m_{Z'} \simeq m_{X^3}$ for tiny $\chi$. 
The mass eigenstates are written by 
\begin{equation}
\begin{pmatrix} Z \\ Z' \end{pmatrix} = \begin{pmatrix} \cos \theta_{ZZ'} & \sin \theta_{ZZ'} \\ - \sin \theta_{ZZ'} & \cos \theta_{ZZ'} \end{pmatrix} \begin{pmatrix} \tilde Z \\ X^3 \end{pmatrix}.
\end{equation}
In Fig.~\ref{fig:mixing}, we show $\sin \theta_{ZZ'}$ as a function of $m_{X^\pm}$ for $\chi =10^{-2}$ and $10^{-3}$ where we fix $2v_\varphi^2/v_\Phi^2 = 0.01$.
We find that $\sin \theta_{ZZ'} \lesssim 10^{-3}$ for $\chi \leq 10^{-2}$ and $m_{X^\pm} \geq 100$ GeV, and thus it is safe from current constraints~\cite{Langacker:2008yv}.

%%%%%%%%%%%%%%%%%%%
%%%%%%%%%%%%%%%%%%%

Finally interactions among dark gauge fields are also given by
\begin{equation}
\mathcal{L} \supset - g_X \epsilon^{abc} \partial_\mu X^a_\nu X^{b \mu} X^{c \nu} - \frac14 g_X^2 \epsilon^{abc} \epsilon^{ade} X^b_\mu X^c_\nu X^{d \mu} X^{e \nu},
\label{eq:gauge_int}
\end{equation}
where $\epsilon^{abc}$ is the structure constants of $SU(2)_X$  and $a = 1,2,3$.
Since $m_{Z'} \simeq m_{X^3} \gtrsim m_{X^\pm}$ the four point gauge interaction in Eq.~(\ref{eq:gauge_int}) does not provide 
dominant contribution to DM annihilation process. 
We thus focus on the three point interaction which is given by $X^\pm$ and $Z'$ such that 
\begin{align}
\mathcal{L} & \supset i g_X C_{ZZ'} \Bigl[ (\partial_\mu X^+_\nu - \partial_\nu X^+_\mu) X^{- \mu} Z'^\nu - (\partial_\mu X^-_\nu - \partial_\nu X^-_\mu) X^{+ \mu} Z'^\nu \nonumber \\
& \qquad \qquad \quad + \frac{1}{2} (\partial_\mu Z'_\nu - \partial_\nu Z'_\mu) (X^{+\mu } X^{- \nu} - X^{- \mu} X^{+ \nu}) \Bigr],
\end{align}
where $C_{ZZ'} \equiv \cos \theta_{ZZ'}$.
Then $Z'$ interacts with the SM particles via $Z$--$Z'$ mixing.
The $Z'$ interaction with SM fermions $f$ is given by 
\begin{equation}
\mathcal{L}_{Z' \bar f f} = \frac{g}{\cos \theta_W} Z'_\mu \bar f \gamma^\mu \left[ - S_{ZZ'} (T_3 - Q \sin^2 \theta_W) + C_{ZZ'} \chi Y  \sin \theta_W  \right]   f,
\end{equation} 
where $T_3$ is diagonal generator of $SU(2)_L$, $Q$ is the electric charge, $S_{ZZ'} \equiv \sin \theta_{ZZ'}$ and $C_{ZZ'} \equiv \cos \theta_{ZZ'}$.
%%%

%The breaking pattern of $\varphi$ is derived by the same way of $\Phi$.
%$\varphi$ is transformed under the unitary transformation of $V=e^{i\pi T_3}$ as follows:
%$V\varphi=[-\varphi_{1,1}^{(3)}, (v_\varphi+\varphi_{1,0}^{(2)})/\sqrt2, -\varphi_{1,-1}^{(1)} ]^T$,
%where $T_3\equiv{\rm diag}(1,0,-1)$; therefore $V\langle\varphi\rangle=V\langle\varphi\rangle$.
%%Then, each of the components of $\varphi$ has parity odd and even; $Parity(\varphi_{1,1}^{(3)}, (v_\varphi+\varphi_{1,0}^{(2)})/\sqrt2, -\varphi_{1,-1}^{(1)})=(-,+,-)$.
%It implies that we also have a remnant $Z_2$ symmetry from $\varphi$.
%\\
%%%%

\subsection{Dark matter physics}

Here we discuss dark matter physics such as relic density and direct detection by carrying out numerical analysis.
In our analysis we write $Z'$ mass by
\begin{align}
m_{Z'} \simeq 2 m_{X^\pm} \left( 1+ R_M \right)^{-\frac12}
\end{align}
where we ignored tiny kinetic mixing effect and $R_M \equiv v_\varphi^2/v_\Phi^2$ representing deviation from the relation $m_{Z'} = 2m_{X^\pm}$.
We expect $R_M << 1$ so that we can explain the sizable relic density by resonant enhancement.
%This is because there is a resonant point under this condition and the cross section is enhanced.
%%%
Note that if the hidden gauge coupling $g_X$ is very large around order $1$, the resonant point would not be needed to explain the correct relic density.
Performing numerical analysis, we search for parameter region satisfying observed relic density of $\Omega h^2 \sim 0.12$~\cite{pdg}.

\begin{figure}[t]
\begin{center}
\includegraphics[width=100mm]{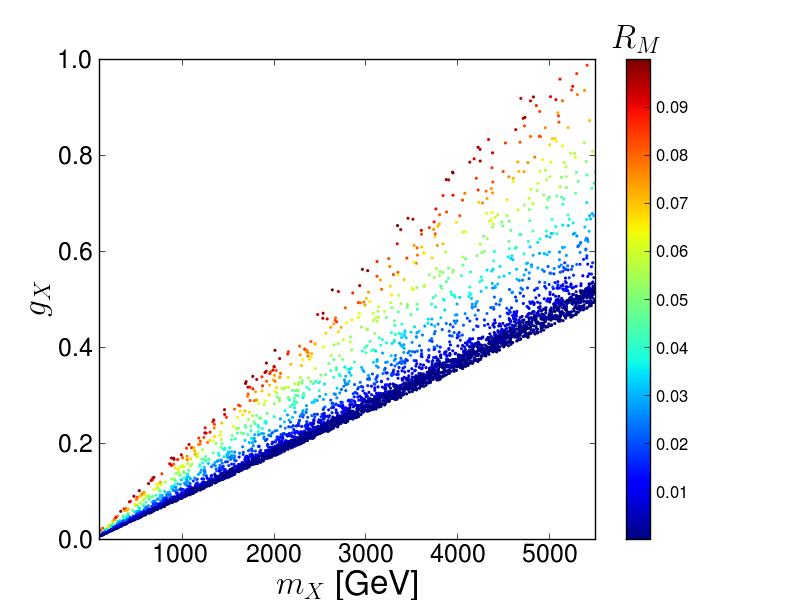} 
\caption{The parameter points which realize relic density of DM as $0.11 \leq \Omega h^2 \leq 0.13$; we fix $\chi = 10^{-3}$ here. 
The color gradient indicates the value of $R_M$. } 
 \label{fig:DM}
\end{center}\end{figure}

We calculate relic density of DM with {\it micrOMEGAs 5}~\cite{Belanger:2014vza} implementing relevant interactions.
In our analysis, we randomly scan parameters $\{m_{X^\pm}, g_X, R_M \}$ within the following region
\begin{equation}
m_{X^\pm} \in [100, 5500] \ {\rm GeV}, \quad g_X \in [0.001, 1.0], \quad R_M \in [0.001, 0.1],
\end{equation}
where we consider two values of kinetic mixing parameter as $\chi = 10^{-2}$ and $10^{-3}$.
Each point in Fig.~\ref{fig:DM} shows parameter sets which realize relic density of DM in approximated observed region of $0.11 \leq \Omega h^2 \leq 0.13$;
here we fix $\chi = 10^{-3}$ but the behavior will be the same if we choose $\chi=10^{-2}$ since only $Z'$ width is affected by the change. 
Here color gradient in the figure indicates the value of $R_M$ as shown in color bar.
We find that observed relic density can be explained by $0.01 \lesssim g_X \lesssim 1.0$ for DM mass range $[100, 5500]$ GeV 
where $g_X$ tends to be larger for large $R_M$ since resonant enhancement becomes smaller.

%%%%%%%%%%%%%%%%%%%
%%%%%%%%%%%%%%%%%%%

\begin{figure}[t]
\begin{center}
\includegraphics[width=60mm]{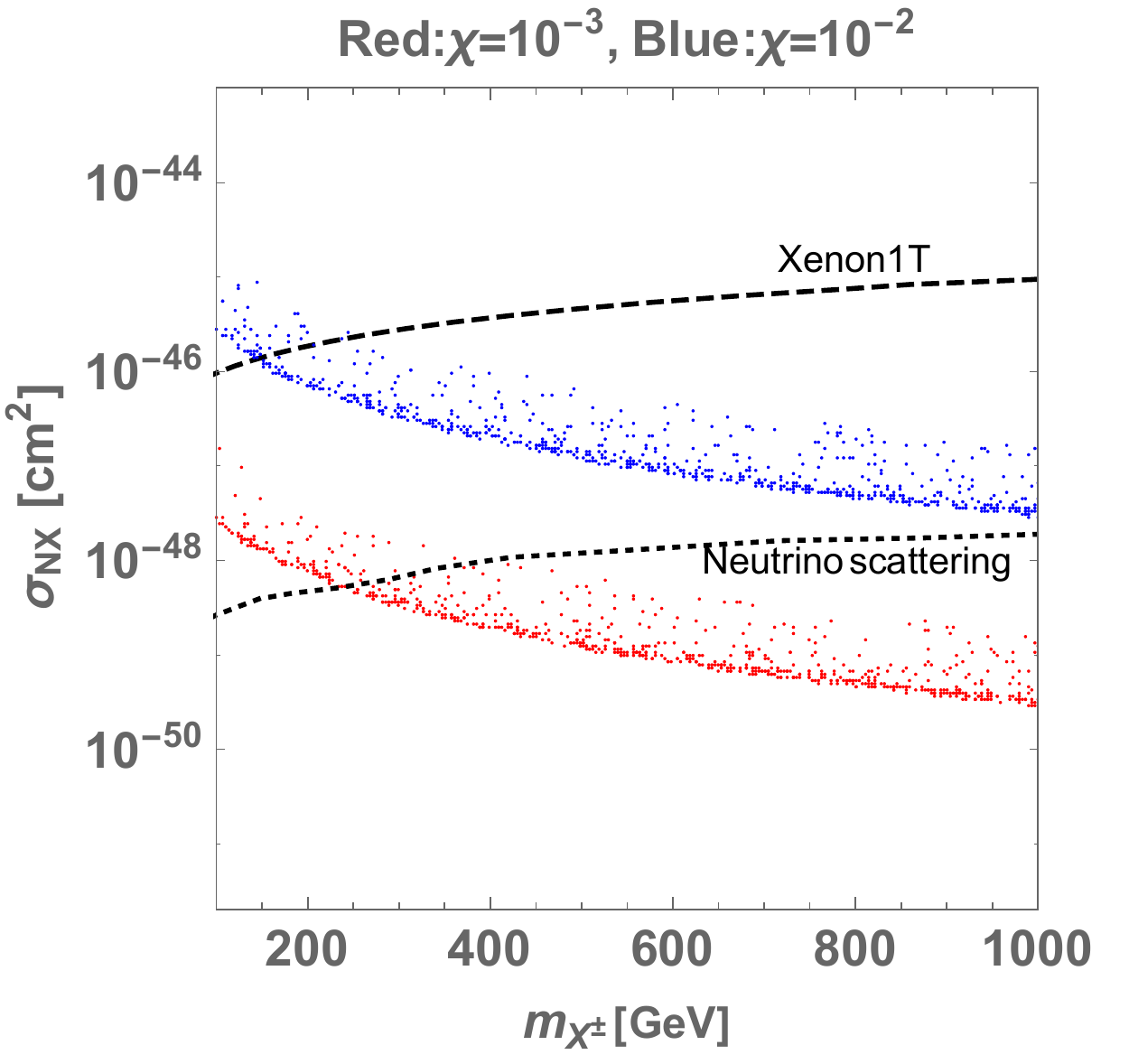}
\includegraphics[width=60mm]{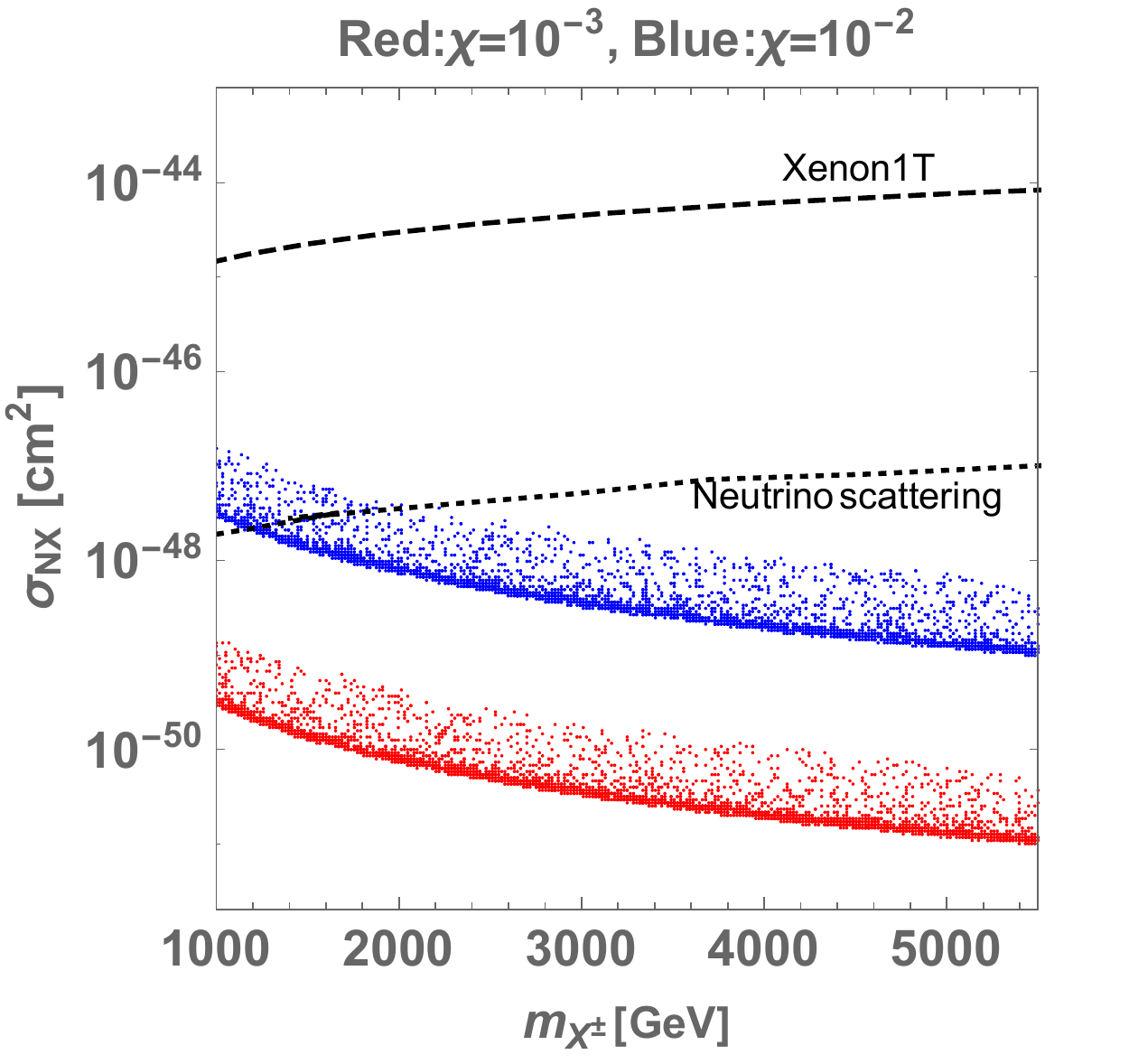} 
\caption{Spin independent DM-nucleon scattering cross section for allowed parameter sets where dashed curve corresponds to XENON1T constraint and dotted curve corresponds to neutrino scattering limit.} 
 \label{fig:DD}
\end{center}\end{figure}

Next we discuss direct detection of DM calculating nucleon-DM scattering cross section for allowed parameter sets which explain observed relic density.
In our case nucleon and DM interact via $Z$ and $Z'$ exchanging process due to $Z$--$Z'$ mixing effect.
Spin independent nucleon-DM scattering cross section is obtained by
\begin{equation}
\sigma_{N X} = \frac{2 g_X^2 g^2}{\pi \cos^2 \theta_W} \left( \frac{m_{X^\pm} m_N}{m_X + m_N} \right)^2 \left( \frac{S_{ZZ'} C^V_{NNZ}}{m_{Z}^2} + \frac{C_{ZZ'} C^V_{NNZ'}}{m_{Z'}^2} \right)^2,
\end{equation}
where $m_{N}$ is nucleon mass ($N = p,n$). 
The nucleon-$Z(Z')$ vector coupling $C^V_{NNZ(Z') }$ is given by 
\begin{align}
& C_{ppZ}^V = C_{ZZ'} \left( \frac14 - \sin^2 \theta_W \right), \quad C_{nnZ}^V = C_{ZZ'} \left(- \frac14 \right), \nonumber \\
& C_{ppZ'}^V = - S_{ZZ'} \left( \frac14 - \sin^2 \theta_W \right) + \frac32 C_{ZZ'} \sin \theta_W \delta, \quad C_{nnZ'}^V =  \frac14 S_{ZZ'} + \frac12 C_{ZZ'} \sin \theta_W \delta.
\end{align}
In Fig.~\ref{fig:DD}, we show $\bar \sigma_{NX} \equiv (\sigma_{pX} + \sigma_{nX})/2$ as a function of DM mass for allowed parameter sets where 
left and right plots show DM mass range $[100, 1000]$ GeV and $[1000, 5500]$ GeV respectively.
We also compare our result with experimental constraints by XENON1T~\cite{Aprile:2017iyp,Aprile:2018dbl} and neutrino scattering limit~\cite{Billard:2013qya}.
For $\chi = 10^{-3}$, the scattering cross sections are much less than current bound. 
They are even smaller than neutrino scattering limit except for light DM region $m_{X^\pm} \lesssim 250$ GeV.  
For $\chi = 10^{-2}$, the scattering cross sections are above current bound in $m_{X^\pm} \lesssim 150$ GeV region, 
between neutrino scattering limit and current bound in  150 GeV $\lesssim m_{X^\pm} \lesssim  1200$ GeV, and below neutrino scattering limit in $m_{X^\pm} \gtrsim  1200$ GeV.
Thus DM signal could be detected for $\chi = 10^{-2}$ and $m_{X^\pm} \lesssim 1200$ GeV in future experiments 
while it is difficult to test DM signal by direct detection for small kinetic mixing parameter $\chi$.

%%%%%%%%%%%%%%%%%%%
%%%%%%%%%%%%%%%%%%%

Before closing this section we comment on DM production at collider experiments.
Our DM candidate $X^\pm$ can be produced at collider via $Z'$ and/or scalar portal 
when we have $Z$--$Z'$ mixing and/or mixing between new scalar from $\Phi^{(')}$ and SM Higgs.
From $Z$--$Z'$ mixing, we would have $q \bar q \to Z' \to X^+ X^-$ process where we need extra jet/photon to detect at collider experiments.
From scalar mixing, we would have $gg \to \phi \to X^+ X^-$ process where $\phi$ is a scalar boson which is mixture of SM Higgs and a new scalar boson. 
Detailed study of collider signal is beyond the scope of this paper and it will be given in future works.

\section{ Conclusions and discussions}
We have proposed a scenario of dark sector described by hidden $SU(2)$ gauge symmetry which is broken by a VEV of scalar multiplet.
Firstly we have discussed general mass relation among $SU(2)$ gauge bosons after spontaneous symmetry breaking 
which is determined by weight of gauge group representation associated with a scalar multiplet developing VEV.  
Interestingly, when $SU(2)$ is broken by scalar quintet VEV, hidden gauge bosons can be DM and mediator with mass ratio of $1:2$ which realizes resonant DM annihilation.
In addition $Z_2$ symmetry stabilizing DM is naturally obtained in this case.  

Then we have discussed a model introducing hidden $SU(2)_X$ quintet and triplet scalar fields in which resonant DM annihilation is realized.
In this model dark sector and the SM sector interact through kinetic mixing between $SU(2)_X$ and $U(1)_Y$ gauge fields obtained from higher dimensional operator through the VEV of $SU(2)_X$ triplet scalar.
We have shown the kinetic mixing term and formulated $Z$--$Z'$ mixing to give interaction between DM and the SM particles.
After formulating interactions, we estimated DM relic density and DM-nucleon scattering cross section.
We have found that DM relic density can be explained for $SU(2)_X$ gauge coupling in the range of $\sim [0.01, 1]$.
In addition, most of parameter sets giving observed relic density is safe from direct detection constraints and some region can be tested in future experiment.

%\section*{ Appendix}
%%%%%%%%%%%%%%%%%%%...

%\newpage
%%%%%%%%%%%%%%%%%%%%%%%%%%%%%%%%%%%
\section*{Acknowledgments}
\vspace{0.5cm}
{\it
This research was supported by an appointment to the JRG Program at the APCTP through the Science and Technology Promotion Fund and Lottery Fund of the Korean Government. This was also supported by the Korean Local Governments - Gyeongsangbuk-do Province and Pohang City (H.O.). H. O. is sincerely grateful for the KIAS member, and log cabin at POSTECH to provide nice space to come up with this project.
The work of S.Y. is supported by the research grant ``The Dark Universe: A Synergic Multi-messenger Approach'' number 2017X7X85K under the program PRIN 2017 funded by the Ministero dell’Istruzione Universit\`a e della Ricerca (MIUR).}
%%%%%%%%%%%%%%%%%%%%%%%%%%%%%%%%%%%
%%%%%%%%%%%%%%%%%%%%%%%%%%%%%%%%%%%

\end{document}